\documentclass[12pt]{article}
\usepackage[dvips]{graphicx}

\begin{document}

\title{Some properties of evolution
equation for
homogeneous
nucleation period under the
smooth behavior of initial conditions}
\author{Victor Kurasov}

\maketitle

\begin{abstract}
The properties of the evolution
equation have been analyzed. The
uniqueness  and the existence of
solution for the evolution equation
with special value of parameter
characterizing intensity of change of
external conditions, of the
corresponding iterated equation have
been established. On the base of these
facts taking into account some
properties of behavior of solution the
uniqueness of the equation appeared in
the theory of homogeneous nucleation
has been established. The equivalence
of auxiliary problem and the real
problem  is shown.
\end{abstract}

\section{Formulation of the problem}

In \cite{PhysRevE94} the evolution
equation for the nucleation period has
been derived. Here we shall analyze the
properties of the evolution equation
and the uniqueness and existence of the
solution of special problems appeared
in the theory of nucleation.

Consider the initial problem
formulated in \cite{PhysRevE94}
{\bf
\begin{itemize}
\item
The following equation is given
\begin{equation} \label{i}
g(z) = f_* \int_{-\infty}^z (z-x)^3
\exp(cx -\Gamma g(x) / \Phi_*) dx
\end{equation}
with positive parameters $f_*$,
$\Gamma$, $\Phi_*$, $c$, which are
chosen to satisfy condition
\begin{equation} \label{ii}
\frac{d g(z)}{dz} |_{z=0} = c \Phi_* /\Gamma
\end{equation}
\end{itemize}
}
Equation (\ref{i}) and condition
(\ref{ii}) form the initial problem.

This problem as well as all other ones is
considered in a class of  continuous
functions.

Instead of the function $g$ we introduce
the function $G = \Gamma g / \Phi_*$.
Then we get
$$
G(z) = ( f_* \Gamma / \Phi_* )
\int_{-\infty}^z (z-x)^3 \exp(cx - G(x))
dx
$$
$$
\frac{d G(z)}{dz} |_{z=0} = c
$$

We change the variables
$$
z \rightarrow ( f_* \Gamma / \Phi_*
)^{-1/4} z'
$$
$$
x \rightarrow ( f_* \Gamma / \Phi_*
)^{-1/4} x'
$$

The new variables will be marked by the
same letters $z,x$.

Then we get
$$
G(z) =
\int_{-\infty}^z (z-x)^3
\exp(c( f_* \Gamma / \Phi_* )^{-1/4}x - G(x))
dx
$$
$$
\frac{d G(z)}{dz} |_{z=0} =
c( f_* \Gamma / \Phi_* )^{-1/4}
$$

We denote $c( f_* \Gamma / \Phi_*
)^{-1/4}$ also as $c$.
The value $G $ will be denoted as  $g$.
We come to the following problem
{\bf
\begin{itemize}
\item
The following equation
\begin{equation}\label{p}
g(z) =  \int_{-\infty}^z (z-x)^3 \exp(cx
- g(x)) dx
\end{equation}
is given with  the positive parameter
$c$, chosen to satisfy
\begin{equation}\label{pp}
\frac{d g(z)}{dz} |_{z=0} = c
\end{equation}
\end{itemize}
}
The equation (\ref{p}) and the condition
(\ref{pp}) form the reduced problem.

So, here only one parameter $c$ remains.

It is more convenient to consider
instead of $g$ the function
$$
\phi = cx -g
$$
Then we come to the following problem
(Problem A)
{\bf
\begin{itemize}
\item
The  following equation
\begin{equation} \label{1}
cz = \phi(z) + \int_{-\infty}^z (z-x)^3
\exp(\phi(x)) dx
\end{equation}
is given with the positive parameter
$c$, chosen to satisfy
\begin{equation} \label{1l}
\frac{d \phi (z)}{dz} |_{z=0} = 0
\end{equation}
\end{itemize}
}
Equation (\ref{1}) and the condition
(\ref{1l}) form the problem A.
This problem will be the auxiliary
problem.

Consider now the problem B
{\bf
\begin{itemize}
\item
The following equation
\begin{equation}\label{g}
3\int_{-\infty}^0 y^2 \exp(\phi(y)) dy z
= \phi(z) + \int_{-\infty}^z (z-x)^3
\exp(\phi(x)) dx
\end{equation}
is given
\end{itemize}
}
Only the equation (\ref{g})
forms  the problem B.

We see that the problem B  has  no
parameters.

The main goal of investigation will be
to see the uniqueness of the solution of
the problem B.

\section{Solution of equation (\ref{1})
with fixed $c$}

\subsection{Some properties of  solution}

At first we consider
 (\ref{1})
 with some positive $c$.

 We rewrite
(\ref{1})  for function $g$  in the form
$$
g(z) =
\int_{-\infty}^z (z-x)^3  \exp(cx-g(x)) dx
$$

One can see the following property

{\bf
If the solution exists, then
$$ g(z) >0 $$
for any $z$
}

This property goes from the positivity of sub-integral
function.

We
introduce the iterated equation
\begin{equation}
\label{3}
g(z) =
\int_{-\infty}^z (z-x)^3  \exp(cx-
\int_{-\infty}^x (x-y)^3  \exp(cy-g(y)) dy) dx
\end{equation}

 We also
introduce the nonlinear operator $P$
according to
$$
P(s) = \int_{-\infty}^z (z-x)^2
\exp(cx-s(x)) dx
$$

One can prove the following fact:

{\bf
If for any $x$ the following inequality
$$ g_1 (x) > g_2(x) $$
takes place, then
$$
P(g_1(x))< P(g_2(x))
$$
takes place for any $x$.
}

It follows  from the explicit form of
$P$.

One can also see the following fact

{\bf
If the solution exists then
$$ g(z) < \exp(cz) /6 $$
for any $z$
}

It follows from the
positivity of exponent in the sub-integral
function.

We construct iterations according to
$$
g_{i+1} = P(g_i)
$$
As the initial approximation we choose
$$
g_0 = 0
$$

{\bf
We see that
$$
g_2 > 0
$$
for any value of the argument
}

 It
follows from the positivity of exponent
and the whole  sub-integral function.

Form the statements formulated above it
 follows the
following chain of inequalities
$$
0=g_0 < g_2< g_4 < .... < g < .... < g_5
< g_3< g_1
$$
for any value of the argument.

So, the odd iterations
converge and the even iterations
converge. So, the solution of the
iterated equation exists. But the
uniqueness isn't yet proven.

For any initial approximation
which is always positive the iterations
with initial  $g_0 = 0$
estimate them from above.

Since the r.h.s. is positive and
solution has to be positive,
 the iterations with any
initial approximation
have to be positive starting
from some number (the asymptotics at
small arguments are written
explicitly).
Correspondingly, later the iterations
will be estimated by the already
constructed iterations. So, the
convergence can be estimated to
convergence of constructed iterations.

Remark: The analogous iterations have
been constructed by F.M.Kuni in 1984,
but for the problem B. For the problem
B the monotonic functions are absent
and it is impossible to prove in such a
simple
way the convergence of iterations
$$
3\int_{-\infty}^0 y^2 \exp(\phi_i(y)) dy
z = \phi_{i+1}(z) + \int_{-\infty}^z
(z-x)^3 \exp(\phi_i(x)) dx
$$
with
$$
\phi_0 = cx
$$

\subsection{Uniqueness of solution
for small arguments}

Now we shall prove
the existence and uniqueness
of non-iterated equation(\ref{1})
(in the class of continuous functions)

It can be done by several ways.
One of the methods
is to introduce the cut-off  from
the side of small $z$,
i.e. to consider only
$z>-a$
with big and positive parameter $a$.
It was done in
\cite{PhysRevE94}.
Here we shall use more rigorous
method.

We shall show that for given $c$
at
$$
z< z_{fin} = \frac{1}{c} \ln [ \frac{
c^4}{6}
(1-\epsilon)]
$$
with some small positive fixed
$\epsilon$ the solution of  (\ref{1})
exists and it is unique.

We construct iterations $g_{i+1} = P(g_i)$.
The starting approximation is not
important.
We note that they do not go out of the
class of positive continuous functions.
Under the arbitrary initial
approximation, all iterations starting
from the first one will be positive.

We have for approximations
$$
g_{i+1} - g_i  =
\int_{-\infty}^z
(x-z)^3
\exp(cx)
(\exp(-g_i) - \exp(-g_{i-1})) dx
$$

Then
$$
|g_{i+1} - g_i| \leq
\int_{-\infty}^z
(x-z)^3
\exp(cx)
|\exp(-g_i) - \exp(-g_{i-1})| dx
$$

It allows to write the inequality for norms
$||...||$ in
the functional space $C [-\infty, \infty]$
$$
||g_{i+1} - g_i|| \leq
\int_{-\infty}^z
(x-z)^3
\exp(cx)
||\exp(-g_i) - \exp(-g_{i-1})|| dx
$$

One can account that for positive
$g_i$ and positive
$g_{i+1}$ the following
inequality
$$
| \exp(g_i) - \exp(g_{i+1} | < |g_i -
g_{i+1} |
$$
takes place.
Having fulfilled the analogous
transformations we come to
$$
||g_{i+1} - g_i|| \leq
\int_{-\infty}^z
(x-z)^3
\exp(cx)
||g_i -g_{i-1}|| dx
$$

We take $   ||g_i -g_{i-1}|| $
out of the integral and get
$$
||g_{i+1} - g_i|| \leq
\int_{-\infty}^z
(x-z)^3
\exp(cx) dx
||g_i -g_{i-1}||
$$

Calculation of integral leads to
$$
||g_{i+1} - g_i|| \leq
\frac{ 6
\exp(cz)}{ c^4}
||g_i -g_{i-1}||
$$

For the mentioned values of $z$
we have
$$
||g_{i+1} - g_i|| \leq ||g_i -g_{i-1}||
(1-\epsilon)
$$

The application of the recurrent
procedure gives
$$
||g_{i+1} - g_i|| \leq (1-\epsilon)^i
const
$$

It is clear that the r.h.s. allows
summation
$$
\sum_{i=0}^{\infty} (1-\epsilon)^i const =
\frac{1}{\epsilon} const
$$

Then this sequence is the Caushi sequence, then
it converges, then the limit of iterations exists
and it is the solution of equation (\ref{1})
with given $c$.
Since the iterations from the first one
are positive, then the convergence
takes place namely to this limit and the
solution is the unique one.

The proof of existence and the
uniqueness of solution
(\ref{1})  with
given $c$ in the region   $z<z_{in}$
is completed.

Now we shall consider solution at
$z>z_{in}$.
At first it will be difficult to
consider all
$z$,
and we shall consider only those
arguments, which correspond to the
formation of essential part of the
droplets. We shall determine them more
accurately.

Remark: We consider here a set of
iterations.
But this construction has no connection
with a real rate of convergence of
iterations with a fixed initial
approximation. To estimate the real rate of
convergence it
 is more correct  to consider at first
two different solutions, then to
construct the iterations based on these
solutions and to prove that they will
come to one limit. Here the iterations
are constructed correctly, but later
the difficulty appears. The difficulty is
 that one can not use in constructions
 for calculations
the limit of previous iterations as the
base for the next type of iterations.
One has to make a more detailed
analysis including the establishing of
the smoothness of the influence of the
difference of the real iteration and
the
limit on the evolution at the further
sub-period.
It can be done but it is connected with
some long formulas. So, we use the
iterations, but keep in mind that
 we have to consider simply
two different solutions. At the further
sub-periods since the uniqueness at
previous sub-period has been
established one can choose the
different solution with the common
part at the previous sub-period.

\subsection{The properties of the function $\psi$
}

At first we shall present some
properties of solution. One can notice
that
\begin{itemize}
\item
The precise solution $g$
increases if $z$
increases
\item
Any iteration $g_i$
increases if $z$
increases
\end{itemize}

This follows from the explicit
expressions.

The mentioned properties
allow to show some properties
for the function $\psi = \exp(cx -
g)$
and iterations $\psi_i = \exp(cx-
g_i(x))$
\begin{itemize}
\item
Function $\psi$
has only one maximum
\item
Function $\psi_i$ for every $i$
has only one maximum
\item
Function $\psi$
is always positive
\item
Function $\psi_i$
for every $i$ is always positive
\item
The maximum of $\psi$
is greater than the maximum of
$\psi_1$
\item
The maximum of
$\psi$
is less than the maximum of
$\psi_2$
\end{itemize}

Beside this one can show that if
$\psi$
really exists, then all possible
solutions lie between $\psi_1$
and $\psi_2$.

We shall prove that $\psi_2$ at big $x$ has to
decrease until zero. It can be seen from
the following estimate. Certainly,
$$
\psi_2 =
\exp(cz - \int_{-\infty}^z (z-x)^3 \psi_1(x)  dx)
$$
for positive $z$
is less than function
$$
\psi_{20} =
\exp(cz - \int_{-\infty}^0 (z-x)^3 \psi_1(x)  dx)
$$
The last function has the evident
asymptote
$$
\psi_{20} =
\exp(cz - z^3 \int_{-\infty}^0  \psi_1 (x)  dx)
$$
which can be calculated exactly with
some positive
$$
N_+ =\int_{-\infty}^0  \psi_1 (x)  dx =
\int_{-\infty}^0
\exp(cx-\frac{6}{c^4} \exp(cx))  dx
$$
or
$$
N_+ = \frac{c^3}{6} [ 1-
\exp(-\frac{6}{c^4})]
$$

So, we see that $\psi_2$ goes to zero
rather fast and for given small positive
fixed value of $\psi_2$ (let it be $l$)
one can see the boundary $z_{fin}$ where
$\psi_2$ will be smaller than $l$.
This completes the proof.

Since one can prove that solutions
$\tilde{\psi}$ of iterated equation
lie below $\psi_2$ the same
conclusions take place for the
iterated equation.

\subsection{The uniqueness of solution for
the intermediate values of argument}

Now we shall prove that for
$z_{in}<z<z_{fin}$ the solution exists
and it is the unique one.

Here we construct the  iterations of the
special type  $1$ (marked by the
subscript  $sp1$ where it is important)
as the following ones:
\begin{itemize}
\item
Before $z<z_{in}$ the initial
approximation $g_0$ is the precise
solution (we already proved that it is
unique).
\item
At $z>z_{in}$ we take $g_0 = g(z_{in})$
\item
The recurrent procedure remains the
previous one.
\end{itemize}

The old iterations will be marked by
$sp0$.

These iterations satisfy the following
properties
\begin{itemize}
\item
All iterations at $z<z_{in}$ coincide
with precise $g$
\item
Since for $z>z_{in}$ we have
$g>g(z_{in})$,  $g_{sp2} >g(z_{in})$ and
the recurrent procedure in the same one,
we can come to the same chains of
inequalities
$$
g_{sp1\ 0} < g_{sp1\ 2} < ... < g < ...<
g_{sp1\ 3}< g_{sp1\ 1}
$$
\end{itemize}

We denote by $z_{max i}$ the maximum
attained by $\psi_i$ and by $z_{max}$
the maximum of precise $\psi$. One can
see that
$$
\frac{dg}{dz} =
3 \int_{-\infty}^z  (z-x)^2 \exp(cx -g) dx
$$
So, then
$$
z_{max\ sp0\ 1 }< z_{max\ sp0\ 2}
$$
$$
z_{max\ sp0\ 1 }< z_{max }
$$
$$
z_{max}< z_{max\ sp0\ 2}
$$
and
$$
z_{max\ sp1\ 1 }< z_{max\ sp1\ 2}
$$
$$
z_{max\ sp1\ 1 }< z_{max }
$$
$$
z_{max}< z_{max\ sp1\ 2}
$$

For the difference $g_{i+1} - g_i$ one
can write
$$
g_{i+1} - g_i =
\int_{-\infty}^z (z-x)^3 \exp(cx)
[\exp(-g_i(x)) - \exp(-g_{i-1}(x))] dx
$$
Then
$$
g_{i+1} - g_i =
\int_{z_{in}}^z (z-x)^3 \exp(cx)
[\exp(-g_i(x)) - \exp(-g_{i-1}(x))] dx
$$
and
$$
|g_{i+1} - g_i | \leq
\int_{z_{in}}^z (z-x)^3 \exp(cx)
|\exp(-g_i(x)) - \exp(-g_{i-1}(x))| dx
$$

Since $g_i>0$, $g_{i-1}>0$ one can write
$$
|g_{i+1} - g_i | \leq
\int_{z_{in}}^z (z-x)^3 \exp(cx)
|g_i(x) - g_{i-1}(x)| dx
$$
and
$$
||g_{i+1} - g_i || \leq
\int_{z_{in}}^z (z-x)^3 \exp(cx) dx
||g_i(x) - g_{i-1}(x)||
$$
where the norm is taken in $C[-\infty, \infty] $.

 Since $z-x < z_{fin} - z_{in}$ one can come to
$$
||g_{i+1} - g_i || \leq
\int_{z_{in}}^z  \exp(cx) dx
||g_i(x) -
g_{i-1}(x)||(z_{fin}-z_{in})^3
$$
Since $\exp(cx) < \exp(cz_{fin})$ we see
that
$$
||g_{i+1} - g_i || \leq
\int_{z_{in}}^z   dx
||g_i(x) -
g_{i-1}(x)||(z_{fin}-z_{in})^3
\exp(cz_{fin})
$$
and having calculated the integral
we get
$$
||g_{i+1} - g_i || \leq (z-z_{in}) ||g_i(x)
- g_{i-1}(x)||(z_{fin}-z_{in})^3
\exp(cz_{fin})
$$

The recurrent application of the last
estimate (with the explicit integration
of
$(z-z_{in})$) leads to
$$
||g_{i+1} - g_i || \leq
\frac{(z-z_{in})^i}{i!}
const (z_{fin}-z_{in})^3
\exp(cz_{fin})
$$
or simply to
$$
||g_{i+1} - g_i ||=
\frac{(z-z_{in})^i}{i!}
const
$$

Summation of the r.h.s. of the last
relation leads to $\exp(z-z_{in})$. Then
this sequence in the Caushy sequence.
Then it must have a limit. This limit
will be the unique one and it is the
unique precise solution of equation
(\ref{1}).

The uniqueness in the global sense can
be proven by approach used in
investigation of the iterated solution.
It has to be repeated every time we
come to analogous situation.

The existence and the uniqueness of
solution of equation (\ref{1}) for $z
\leq z_{fin}$ are proven.

Generally speaking this proof is
sufficient for uniqueness at every
finite $z$ but here we have estimated
$\exp(cz)$ by $\exp(cz_{fin}$. It is
possible to give more precise estimate,
which will be done  below.

\subsection{The uniqueness for the big
values of argument}

Now we shall prove the existence and
uniqueness for the rest $z$.

We construct  iterations of the
special type $2$. The procedure is
absolutely analogous to the special type
$1$ but now the iterations are the precise
solution until $z_{fin}$.  All
properties mentioned for the special
type $1$ remain here.

Now we rewrite equation for $\phi_i$ and
have
$$
-\phi_{i+1} + \phi_i =
\int_{z_{fin}}^z (z-x)^{\alpha}
[\exp(\phi_i ) - \exp(\phi_{i-1})] dx
$$
Here $\alpha$ is the power $3$. Now we
keep parameter $\alpha$ because in
further investigations \cite{PhysicaA96}
it will be
necessary to prove all for the
arbitrary positive (and not integer)
power.

For the integer power the task is
more simple because having
differentiated several times  we can kill the
integral term and reduce the equation to
the differential equation. Then we can
use all standard theorems for
differential equations.

Then
$$
|-\phi_{i+1} + \phi_i| \leq
\int_{z_{fin}}^z (z-x)^{\alpha}
|\exp(\phi_i ) - \exp(\phi_{i-1})| dx
$$
Since at $z>z_{fin}$ $\phi_i < \phi_2
\leq
max \phi_2 = l$ one can write
$$
|\exp(\phi_i ) - \exp(\phi_{i-1})|<
 r |\phi_i  - \phi_{i-1}|
$$
with some fixed $r$
Then
$$
|-\phi_{i+1} + \phi_i| <
\int_{z_{fin}}^z (z-x)^{\alpha}
|\phi_i  - \phi_{i-1}| r dx
$$
and for the norms in $C$
$$
||-\phi_{i+1} + \phi_i|| <
\int_{z_{fin}}^z (z-x)^{\alpha} dx
||\phi_i  - \phi_{i-1}|| r
$$

Since $(z-x)<(z-z_{fin})$ one can see
that
$$
||-\phi_{i+1} + \phi_i|| <
\int_{z_{fin}}^z dx (z-z_{fin})^{\alpha}
||\phi_i  - \phi_{i-1}|| r
$$

Having calculated the integral we can
finally come to
$$
||-\phi_{i+1} + \phi_i|| <
(z-z_{fin})^{\alpha+1} ||\phi_i  -
\phi_{i-1}|| r
$$

Having applied this estimate
$i$ times with explicit integration
 one can come
to the following estimate
$$
||\phi_{i+1} - \phi_i ||
 <
 const
 \frac{(z-z_{fin})^{w(i)}}{v(i)}
 $$
 with some constant and two functions $v$ and $w$.
These functions have properties
$$
v(i) > i!
$$
$$
w(i) < ([\alpha]+1)i + const_1
$$
where $[\alpha]$ is the minimal integer
number greater than $\alpha$ and
$const_1$ does not depend on $i$

So, we can come to
$$
||\phi_{i+1} - \phi_i ||
 <
 const
 \frac{(z-z_{fin})^{w(i)}}{i!}
 $$

Consider now $z$ which satisfies condition
$$
z-z_{fin} > 1
$$

Then
$$
||\phi_{i+1} - \phi_i ||
 <
 const
 \frac{(z-z_{fin})^{([\alpha]+1)i+const_1}}{i!}
 $$

One can easily see that the term in the
r.h.s. of the previous relation is the
term in Taylor's serial for the function
$$
const (z-z_{fin})^{const_1}
\exp((z-z_{fin})^{[\alpha]+1})
$$

So, it is the Caushy sequence. So, the
initial sequence is also the Caushy
sequence. So, it converges and converges
to the unique solution. So, the
existence and the uniqueness of
solution in the region $z>z_{fin} +1$ is
proven.

One has also to note that in
consideration of the second region there
was absolutely no difference to prove the
property until $z_{fin}$ or until
$z_{fin}+1$. So, in the rest region
$z_{fin} < z < z_{fin} +1$ the existence
and the uniqueness also take place.

\section{Existence of the root}

\subsection{Some estimates
for positions of the maximum of
spectrum}

Now we shall give some estimates for
position of maximum of $\psi$ at small
and big values of $c$.

Let $z_{max}$ be the point (for given $c$
there will
only one point) of the maximum
of precise $\psi$. Let
$z_{max i}$ be the maximum of $\psi_i$.

One can easily see that
$$
g_1 (0) = \frac{6}{c^4}
$$
Then the value $z_{max1}$ can be found
from the maximum of $\exp(cx -
\frac{6}{c^4} \exp(cx))$ and satisfies
the
equation
$$
c = \frac{6}{c^3} \exp(cz_{max1})
$$
Then
$$
z_{max1} = \frac{1}{c} \ln\frac{c^4}{6}
$$
For every positive $c$ it exists.

The value of $\psi_1$ at the maximum is
$$
\psi_{max1} = \exp(cz_{max1} -
\frac{6}{c^4} \exp(cz_{max1})
$$
and
$$
\ln \psi_{max1} = \ln\frac{c^4}{6} -1
$$

The last value goes to infinity when $c$
goes to infinity.

One can see that according to
$$
g'(z) = 3 \int_{-\infty}^z (z-x)^2
\exp(cx - g(x)) dx
$$
the following inequality
$$
g'(z) < g_1'(z)
$$
takes place. One can easily calculate
$g_1'$:
$$
g_1'(z) = 3 \int_{-\infty}^z (z-x)^2
\exp(cx) dx = \frac{6}{c^3}
\exp(cz)
$$

From $ g'(z) < g_1'(z) $ it follows
that
$$
z_{max} > z_{max1}
$$

At $c=6$ we see that
$$
z_{max1} = \frac{1}{6} \ln 6^3  >0
$$

So, at $c=6$ we see that $z_{max} >0$.

Since at every $c$ the solution exists,
since it is unique and since has one maximum
then one can
say that the dependence $z_{max}$ on $c$
can be treated as a function. At $c=6$
it is positive.

\subsection{The case of the small $c$}

Now we shall consider small positive
$c$. We don't mention every time that
$c$ is positive but all conclusions for
arbitrary $c$ mean for arbitrary
positive $c$.

Since
$$
g_2' <g
$$
one can see that
$$
z_{max2} > z_{max}
$$
So, now we calculate $z_{max2}$. It is
the root of equation
$$
c = 3 \int_{-\infty}^{z_{max2}}
(z_{max2}-x)^2
\exp(cx - \frac{6}{c^4} \exp(cx)) dx
$$

We introduce
$$
I_2 (z)  =3 \int_{-\infty}^{z} (z-x)^2
\exp(cx - \frac{6}{c^4} \exp(cx)) dx
$$
as a function of $z$. We see that $I_2$
is a growing function of $z$. The last
equation can be rewritten as
$$
c = I_2
$$
If we change $I_2$ by some $I_2^*$ which
is less than $I_2$
then the root $z_{max2e}$ of
equation
$$
c= I_2^*
$$
will be greater than $z_{max2}$
$$
z_{max2e} > z_{max2}
$$
and, thus,
$$
z_{max2e} > z_{max}
$$

To construct $I_2^*$ we notice that
$$
\exp(-\frac{6}{c^4} \exp(cx))
 >
 1 - \frac{6}{c^4} \exp(cx)
 $$
 since $d^2 \exp(x) / dx^2 >0 $ for all $x$.

Then
$$
I_2 = 3 \int_{-\infty}^z (z-x)^2
\exp(cx)
\exp(-\frac{6}{c^4} \exp(cx)) dx
$$
$$
> 3 \int_{-\infty}^z (z-x)^2 \exp(cx) [1
- \frac{6}{c^4} \exp(cx)] dx = I_2^*
$$

Now we shall calculate the root of
$I_2^* - c$. Having calculated integrals
we come to
$$
c = \frac{6}{c^3} \exp(cx)
- \frac{6}{8c^3} exp(2cx) \frac{6}{c^4}
$$

Having marked $\exp(cx)$ as $y$ we come
to the square algebraic equation
$$
c^4 = 6 y - \frac{36}{8 c^4} y^2
$$

One can see that for small positive $c$
there are two roots: one is very close
to zero, another is greater than the
first one but also it is rather close to zero.
So,
$$
y \approx 0
$$
Then
$$
\exp(cx) \approx 0
$$
and
$
cx $ goes to $-\infty$. It means that
$z \leq z_{max2e}$ goes also to $-\infty$.
Since the roots to equation considered
above are continuous functions of
parameters (at positive $c$) we see that
there exists some fixed small $c_y$  at
which $z_{max2e}$ is negative. Then at
$c_y$ the value $z_{max}$ is also
negative.

If we now prove that the function
$z_{max} (c)$ is a continuous function
then according to Bolzano-Caushi theorem
there will be a root $z_{max} =0$.

\section{The continuous character of
the  dependence of $g$ on $c$}

\subsection{The continuous character of
the  dependence of $g$ on $c$
at moderate arguments}

Now we shall prove that the solution of
equation (\ref{1}) depends on $c$
continuously for $c_y < c < 6$. Then it
will follow that the maximum of $cx-g$
will be continuous function and
$z_{max}$ is continuous function of $c$.

Imagine that $g$ corresponds to $c$ and
$g`$ corresponds to $c`$. Then
$$
g - g`  =
\int_{-\infty}^z (z-x)^3 [\exp(cx)
\exp(-g(x)) - \exp(c`)\exp(-g`(x))]
dx
$$

We see that the function
$$
q  (c,g) = \exp(cx - g(x))
$$
at every $x$ can be interpreted as a
function of two arguments $c$ and $g$.
We see that the dependence on both
arguments is the exponential one. Since
the second derivative of exponent is
always positive, one can give the
estimate from above for $q(c,g) -
q(c`,g`)$. Namely,
$$
|q(c,g) - q(c`,g`)| \leq |(\frac{\partial
q}{\partial c})_m | |c-c`| +
|(\frac{\partial q}{\partial g})_m | |g-g`|
$$
Here index $m$ marks the maximal
absolute values
of derivatives which are attained on one
of the ends of intervals $[c,c`]$ and
$[g,g`]$ correspondingly.

Having calculated derivatives we come to
$$
|q(c,g) - q(c`,g`)| <
\exp(c^* x)
\exp(-g^*(x)) |x|
|c-c`| +
\exp(c^{**} x) \exp(-g^{**} (x))
|g-g`|
$$
Here $c^*$ and $c^{**}$ are some values
of $c$ used in the maximal values of
derivatives; $g^*$ and $g^{**}$ are some
values of $g$ used in the maximal values
of derivatives. We don't require that
$g^*
= g(c^*)$ and $g^{**} = g(c^{**})$.

Then
$$
|g-g`| \leq
\int_{-\infty}^z
(z-x)^3
[
\exp(c^* x)
\exp(-g^*(x)) |x|
|c-c`| +
$$
$$
\exp(c^{**} x) \exp(-g^{**} (x))
|g-g`| ]dx
$$

Having noticed that $\exp(-g^*)<1$ and
$\exp(-g^{**}) <1$ one can come to
$$
|g-g`| \leq
\int_{-\infty}^z
(z-x)^3
[
\exp(c^* x)
 |x|
|c-c`| +
\exp(c^{**} x)
|g-g`| ]dx
$$

Then for the norm in $C$ we have
$$
||g-g`|| \leq
\int_{-\infty}^z
(z-x)^3
[
\exp(c^* x)
 |x|
|c-c`| +
\exp(c^{**} x)
||g-g`|| ] dx
$$
or
$$
|| g-g`|| \leq |c-c`| A + B||g-g`||
$$
where
$$
0 < A = \int_{-\infty}^z (z-x)^3 |x| \exp(c^*
x) dx < \infty
$$
$$
B = \int_{-\infty}^z (z-x)^3
\exp(c^{**} x) dx
$$

The function $B(z)$ is the growing
function of $z$. We see that $B= 6
\exp(c^{**}z) /(c^{**4})$. The condition
$$
B = 1- \epsilon_1
$$
with a fixed small positive $\epsilon_1$
determines the boundary $z_{in*}$. then
for $z<z_{in*}$ we see that
$$
|| g-g`|| < \frac{1}{\epsilon_1} |c-c'|
const
$$
with some fixed $const$. So, $g$ depends
on $c$ continuously at $z<z_{in*}$.

One can also prove that
$$
I_i =
\int_{-\infty}^{z_{in*}} x^i
\exp(cx) \exp(-g(x)) dx
\ \ \ \
i=0,1,2,3
$$
depend on $c$ continuously.

It can be done quite analogously.
Certainly, the value of $z_{in*}$ can be
changed but remains the finite one.

\subsection{The continuous character of
the  dependence of $g$ on $c$ at big arguments}

Now we shall investigate $z>z_{in*}$.
One can present $g-g`$ in the following
form
$$
|g-g`| = T_1 + T_2
$$
$$
T_1 = \int_{-\infty}^{z_{in*}} (z-x)^3
|\exp(cx -g(x)) - \exp(c`x - g`(x)) | dx
$$
$$
T_2 = \int_{z_{in*}}^z (z-x)^3 |\exp(cx
-g(x)) - \exp(c`x - g`(x)) | dx
$$

It is clear that $T_1$ is the polynomial
of the third order on $z$ with coefficients
proportional to $I_i$. Since
coefficients are continuous functions of
$c$ one can state that the whole
polynomial is the continuous function of
$c$. So, one can write
$$
T_1 \leq \alpha_1 |c-c`|
$$
with some constant $\alpha_1$.

Then for $T_2$ one can see that
$$
T_2 \leq
\int_{z_{in*}}^z (z-x)^3
[\exp(c^* x) \exp(-g^*) |x| |c-c`| +
$$
$$
\exp(c^{**} x) \exp(-g^{**}(x))
|g(x)-g`(x)| ] dx
$$

Then since $ \exp(-g^*) < 1$, $
\exp(-g^{**}) < 1$ and $\exp(c^* x) \leq
\exp(c^* z)$, $\exp(c^{**} x) \leq
\exp(c^{**} z)$ one can come to
$$
T_2 \leq
\int_{z_{in*}}^z (z-x)^3
[\exp(c^* z) |x| |c-c`| +
\exp(c^{**} z)
|g(x)-g`(x)| ] dx
$$

Then since $(z-x)^3 < (z-z_{in*})^3$ one
can come to
$$
T_2 \leq T_{21} +T_{22}
$$
$$
T_{21} =
\int_{z_{in*}}^z (z-z_{in*})^3
|z|
\exp(c^* z) dx
|c-c`|
$$
$$
T_{22} =
\int_{z_{in*}}^z (z-z_{in*})^3
\exp(c^{**} z)
|g(x)-g`(x)| dx
$$
Expression for $T_{21}$ and $T_{22}$
can be easy calculated analytically.

When we shall write the estimate for the
norm in $C$. We simply have to write the
norm $|| g-g`||$ instead of the
absolute value $| g-g`|$ and then we
can move $||g-g`||$ out of integral

Then
$$
|| g -g`||  \leq T_1 + T_{21} + T`_{22}
$$
where
$$
T`_{22} =
\int_{z_{in*}}^z (z-z_{in*})^3
\exp(c^{**} z) dx
||g(x)-g`(x)||
$$

The integral in expression for $T`_{22}$
can be taken and
$$
T`_{22} =(z-z_{in*})^4
\exp(c^{**} z)
||g(x)-g`(x)||/4
$$

Both $T_1$ and $T_{21}$ have the form
$$
const |c-c`|
$$ which is necessary and only
$T`_{22}$ is expressed through
$||g(x)-g`(x)||$. So, to have the
convergence after the recurrent use of
the last formula one has to require
$$
(z-z_{in*})^4
\exp(c^{**} z) /4 < 1 - \epsilon_2
$$
with a small fixed positive
$\epsilon_2$.

This completes the procedure.

We can
fulfill this step again and again and
$z$ will grow.

For any arbitrary fixed $z_{fin*}$ we
see that $\exp(c^{**} z) <
\exp(c^{**} z_{fin*})$
and we have
$$
(z-z_{in*}) <
(\frac{4(1-\epsilon_2)}{\exp(c^{**}
z_{fin*})})^{1/4}
$$
So, the size of the step at elementary
procedure will be
$$
\Delta z < (\frac{4(1-\epsilon_2)}{\exp(c^{**}
z_{fin*})})^{1/4}
$$
Since $\Delta z $ is finite, one can
attain $z_{fin}$ by the final number of
steps.

As the result the continuous character
of dependence of $g$ on $c$ is proven.

\subsection{The computational
reasons for the uniqueness of the root of
$z_{max}(c)$}

Since the spectrum $\phi$ has only one
maximum, the continuous character of
dependence of $z_{max}$ on $c$ is
proven.

Since $z_{max}(c_y)<0$ and $z_{max}(c=6)
>0$, the continuous function
$z_{max}(c)$ must have a root. This root
is a solution of the problem A. Thus,
the
existence of solution of the problem A
is proven.

But the uniqueness isn't proven. To do
this we simply calculate $z_{max}$ as a
function of $c$. This dependence is
drawn in the Figure 1.

\begin{figure}[hgh]

\includegraphics[angle=270,totalheight=8cm]{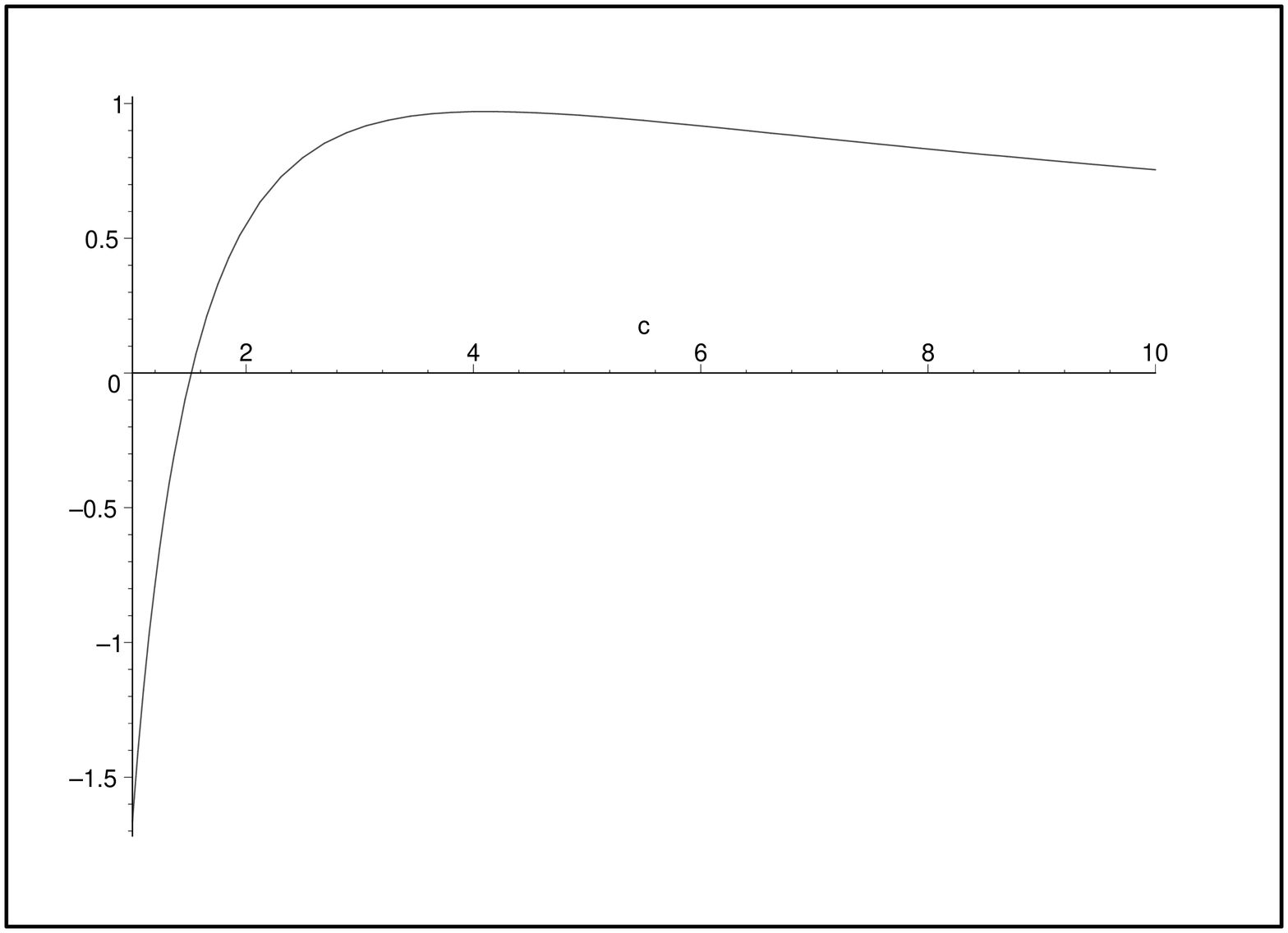}

\begin{caption}
{  Coordinate $z_{max} $ as a function of $c$.}
\end{caption}

\end{figure}

One can see that the root is the unique
one. The analytical proof can be found
below but it requires another
representation.

One can see that for $c<3$ the
dependence  $z_{max}(c)$ is a growing
function. But for $c>(6\exp(1))^{1/4}$
already the first iteration for spectrum
has positive maximum. Since the maximum
of precise solution can not be lower
than the maximum of precise solution.
The maximum of precise solution has
to be positive. But the precise solution
can not lie higher  than $cx$. So, it
must have maximum at positive $x$. It
means that the slow decrease of
$z_{max}(c)$ which is seen in the
figure, can not lead to the negative
$z_{max}$.

So, the solution of the problem A is
also the unique one.

\section{Iterated equation}

\subsection{Some properties  of solution of the
iterated equation}

For further purposes it will be
necessary to investigate the iterated
equation and to show the properties of
solution.
The iterated equation will be written
in the following form
\begin{equation} \label{it}
\tilde{g}(z) =
\int_{-\infty}^z (z-x)^3
\exp(cx - \int_{-\infty}^x (x-y)^3
\exp(cy - \tilde{g}(y)) dy ) dx
\end{equation}

One can see the following property

{
\bf
The solution of (\ref{it}) is so, that
the spectrum, i.e. the function
$$\tilde{\psi} = \exp(cx - \tilde{g})$$
has only one maximum.
}

Really, $\tilde{g}$ is the integral
with the positive sub-integral
function. The function $d\tilde{g} /
dz$
can be presented as
$$
\frac{d\tilde{g}(z)}{dz} =
3 \int_{-\infty}^z (z-x)^2
\exp(cx - \int_{-\infty}^x (x-y)^3
\exp(cy - \tilde{g}(y)) dy ) dx
$$
It is also the integral with the
positive sub-integral function. Then
the function $d\tilde{g} /
dz$ is also positive.

Now we recall functions
$$
g_0 = 0
$$
$$
g_1 = \frac{6}{c^4} \exp(cx)
$$
$$
g_2 = \int_{-\infty}^z (z-x)^3 \exp(cx
-\frac{6}{c^4} \exp(cx) ) dx
$$
They are simply the already
constructed iterations for
eq. (\ref{1}).

One can easily see that every solution
of (\ref{it}) has to satisfy
$$
g_2 \leq \tilde{g} \leq g_1
$$

The proof is simply the comparison
between expressions for $g_1$, $g_2$
and the iterated equation.

\subsection{The uniqueness of the
solution of the iterated equation at
small $z$}

Here one has to fulfill two procedures:
\begin{enumerate}
\item
To construct  iterations to see the
existence of solution
\item
To construct estimates for different
solutions to see the uniqueness
\end{enumerate}
These  procedures have similar
technical realization. This is the
reason why here we  restrict the
consideration  only by the second case.
The first case is identical in ideology
 to the procedure used for original
 (not iterated) equation. There we
 solved only the first problem. The
 second case for the original (non
 iterated) solution can be solved as it
 is described here.

Actually the existence for the iterated
solution is already proven as the limit
of increasing sequence  (odd
iterations) restricted from above and
the limit of decreasing sequence (even
iterations) restricted from below.

The proof is the following:

Suppose that $\tilde{g}$ and $\tilde{q}$ are two
different solutions of equation
(\ref{it}). Then
$$
\tilde{g}(z) - \tilde{q}(z)=
\int_{-\infty}^z (z-x)^3
[
\exp(cx - \int_{-\infty}^x (x-y)^3
$$
$$
\exp(cy - \tilde{g}(y)) dy )
-
\exp(cx - \int_{-\infty}^x (x-y)^3
\exp(cy - \tilde{q}(y)) dy )]
dx
$$

We shall formally denote the initial
functions as $\tilde{g}_0$ and $\tilde{q}_0$
and determine the "formal iterations"
$\tilde{g}_i$ and $\tilde{q}_i$
as
$$
\tilde{g}_{i+1}(z) =
\int_{-\infty}^z (z-x)^3
\exp(cx - \int_{-\infty}^x (x-y)^3
\exp(cy - \tilde{g}_{i}(y)) dy ) dx
$$
$$
\tilde{q}_{i+1}(z) =
\int_{-\infty}^z (z-x)^3
\exp(cx - \int_{-\infty}^x (x-y)^3
\exp(cy - \tilde{q}_{i}(y)) dy ) dx
$$
Then we have
$$
\tilde{g}_{i+1}(z) - \tilde{q}_{i+1}(z)=
\int_{-\infty}^z (z-x)^3
[
\exp(cx - \int_{-\infty}^x (x-y)^3
$$
$$
\exp(cy - \tilde{g}_i(y)) dy )
-
\exp(cx - \int_{-\infty}^x (x-y)^3
\exp(cy - \tilde{q}_i(y)) dy )]
dx
$$

We consider
$
z \leq z_{in}$. Recall that $z_{in}$ is
chosen from condition
$$
\frac{6}{c^4} \exp(cz_{in}) = 1 -
\epsilon
$$
with a small positive $\epsilon$.

We see that
$$
z_{in} \leq z \leq x \leq y
$$

One can reorganize the previous
equation as
$$
\tilde{g}_{i+1}(z) - \tilde{q}_{i+1}(z)=
\int_{-\infty}^z (z-x)^3 \exp(cx)
[
\exp( - \int_{-\infty}^x (x-y)^3
\exp(cy - \tilde{g}_i(y)) dy )
-
$$
$$
\exp( - \int_{-\infty}^x (x-y)^3
\exp(cy - \tilde{q}_i(y)) dy )]
dx
$$

Then
$$
|\tilde{g}_{i+1}(z) - \tilde{q}_{i+1}(z)|
\leq
\int_{-\infty}^z (z-x)^3 \exp(cx)
|
\exp( - \int_{-\infty}^x (x-y)^3
\exp(cy - \tilde{g}_i(y)) dy )
-
$$
$$
\exp( - \int_{-\infty}^x (x-y)^3
\exp(cy - \tilde{q}_i(y)) dy )|
dx
$$

Since for positive $\alpha_1$,
$\alpha_2$ one can see that
$$
|\exp(-\alpha_1) -\exp(-\alpha_2) |
\leq
|\alpha_1 - \alpha_2 |
$$
and
$\int_{-\infty}^x (x-y)^3
\exp(cy - \tilde{g}_i(y)) dy$,
$
\int_{-\infty}^x (x-y)^3
\exp(cy - \tilde{q}_i(y)) dy
$
are evidently positive
then
$$
|
\exp( - \int_{-\infty}^x (x-y)^3
\exp(cy - \tilde{g}_i(y)) dy )
-
\exp( - \int_{-\infty}^x (x-y)^3
\exp(cy - \tilde{q}_i(y)) dy )|
\leq
$$
$$
|
 \int_{-\infty}^x (x-y)^3
\exp(cy - \tilde{g}_i(y)) dy )
-
 \int_{-\infty}^x (x-y)^3
\exp(cy - \tilde{q}_i(y)) dy )|
$$

Having rearranged the r.h.s. we come to
$$
|
\exp( - \int_{-\infty}^x (x-y)^3
\exp(cy - \tilde{g}_i(y)) dy )
-
\exp( - \int_{-\infty}^x (x-y)^3
\exp(cy - \tilde{q}_i(y)) dy )|
\leq
$$
$$
 \int_{-\infty}^x (x-y)^3
 \exp(cy)
|\exp( - \tilde{g}_i(y))
-
\exp( - \tilde{q}_i(y))| dy
$$

Since $\tilde{q}_i(y) \geq 0$,
$\tilde{q}_i(y) \geq 0$, one can see
that
$$
|\exp( - \tilde{g}_i(y)) dy )
-
\exp( - \tilde{q}_i(y)) dy )|
\leq
$$
$$
| \tilde{g}_i(y)
-
\tilde{q}_i(y)|
$$
and then
$$
|
\exp( - \int_{-\infty}^x (x-y)^3
\exp(cy - \tilde{g}_i(y)) dy )
-
\exp( - \int_{-\infty}^x (x-y)^3
\exp(cy - \tilde{q}_i(y)) dy )|
\leq
$$
$$
 \int_{-\infty}^x (x-y)^3
\exp(cy) | \tilde{g}_i(y)
-
\tilde{q}_i(y)|dy
$$
Then for the norm in $C$ space one can
write
$$
|
\exp( - \int_{-\infty}^x (x-y)^3
\exp(cy - \tilde{g}_i(y)) dy )
-
\exp( - \int_{-\infty}^x (x-y)^3
\exp(cy - \tilde{q}_i(y)) dy )|
\leq
$$
$$
 \int_{-\infty}^x (x-y)^3
\exp(cy) dy || \tilde{g}_i(y)
-
\tilde{q}_i(y)||
$$
Having calculated the integral one comes
to
$$
|
\exp( - \int_{-\infty}^x (x-y)^3
\exp(cy - \tilde{g}_i(y)) dy )
-
\exp( - \int_{-\infty}^x (x-y)^3
\exp(cy - \tilde{q}_i(y)) dy )|
\leq
$$
$$
 (1-\epsilon) || \tilde{g}_i(y)
-
\tilde{q}_i(y)||
$$

Now we shall estimate
$|\tilde{g}_{i+1} -\tilde{q}_{i+1}|$.
We have
$$|\tilde{g}_{i+1} -\tilde{q}_{i+1}|
\leq
\int_{-\infty}^z (z-x)^3 \exp(cx)
(1-\epsilon) || \tilde{g}_i(y)
-
\tilde{q}_i(y)|| dx
$$
and then
$$|\tilde{g}_{i+1} -\tilde{q}_{i+1}|
\leq
(1-\epsilon)^2 || \tilde{g}_i(y)
-
\tilde{q}_i(y)||
$$

We can apply this
recurrent procedure
several times and come to
$$|\tilde{g}_{i+1} -\tilde{q}_{i+1}|
\leq
(1-\epsilon)^{2i} || \tilde{g}_0(y)
-
\tilde{q}_0(y)||
$$
So, we see that
$|\tilde{g}_{i+1} -\tilde{q}_{i+1}|$
can be estimated by some positive
members of geometric progression. So, it
means that this sequence converges and
the limit is the real unique solution.
The difference between $\tilde{g}_i$ and
$\tilde{q}_i$ can be made negligibly
small.

So, we conclude that $\tilde{g}$ and
$\tilde{q}$ is actually one solution.
The uniqueness is proven. The existence
can be proven by construction of Caushi
sequence.

\subsection{Further properties of solution
}

We need some estimates for
$\tilde{\psi} = \exp(cx - \tilde{g})$.

Having constructed
$$
\tilde{\xi}_i = cx - \tilde{g}_i
$$
$$
\tilde{\xi} = cx - \tilde{g}
$$
we see that
$$
cx - g_1 \leq \tilde{\xi}_i
\leq
cx - g_2
$$
$$
cx - g_1 \leq \tilde{\xi}
\leq
cx - g_2
$$

Then it follows that
$$
max \tilde{\xi}_i \leq
max(cx - g_2)
$$
$$
max \tilde{\xi}_i \geq
max(cx - g_1)
$$
$$
max \tilde{\xi} \leq
max(cx - g_2)
$$
$$
max \tilde{\xi} \geq
max(cx - g_1)
$$
and we have the estimates for the
maximum.

We introduce $z_{21}$ as
to satisfy
$$
(cx - g_2)|_{z_{21}}
 =
 max\{ cx - g_1\}
$$

There are two $z_{21}$. We shall note
them
$z_{21l}$ and $z_{21r}$ and choose to
have
$z_{21l} < z_{21r}$.

One can easily see that
$$
z_{21l} <
\tilde{z}_{max\ i} < z_{21r}
$$
where
$\tilde{z}_{max\ i}$
is the point where the maximum of
$\tilde{\xi}_i$ is attained.
It is also clear that
$$
z_{@1l} <
\tilde{z}_{max} < z_{21r}
$$
where
$\tilde{z}_{max}$
is the point where the maximum of
$\tilde{\xi}$ is attained.

So, now the natural boundary $z_{21r}$
appeared and we shall consider $z<
z_{21r}$.

Since to solve the problem of existence
we have to construct iterations then
the estimates for iterations are
presented.

\subsection{Uniqueness for intermediate
$z$}

Suppose that there are two different
solutions $\tilde{g}$ and $\tilde{q}$. According to the
previous constructions they must coincide
at $z<z_{in}$. Then
$$
\tilde{g}(z) - \tilde{q}(z)=
\int_{z_{in}}^z (z-x)^3
[
\exp(cx - \int_{-\infty}^x (x-y)^3
$$
$$
\exp(cy - \tilde{g}(y)) dy )
-
\exp(cx - \int_{-\infty}^x (x-y)^3
\exp(cy - \tilde{q}(y)) dy )]
dx
$$
Then for the absolute values
$$
|\tilde{g}(z) - \tilde{q}(z) |
\leq
\int_{z_{in}}^z (z-x)^3
|
\exp(cx - \int_{-\infty}^x (x-y)^3
$$
$$
\exp(cy - \tilde{g}(y)) dy )
-
\exp(cx - \int_{-\infty}^x (x-y)^3
\exp(cy - \tilde{q}(y)) dy )|
dx
$$
Then since $z-x \leq z-z_{in}$ we come
to
$$
|\tilde{g}(z) - \tilde{q}(z) |
\leq (z-z_{in})^3
\int_{z_{in}}^z
|
\exp(cx - \int_{-\infty}^x (x-y)^3
$$
$$
\exp(cy - \tilde{g}(y)) dy )
-
\exp(cx - \int_{-\infty}^x (x-y)^3
\exp(cy - \tilde{q}(y)) dy )|
dx
$$

The function $$|
\exp(cx - \int_{-\infty}^x (x-y)^3
\exp(cy - \tilde{g}(y)) dy )
-
\exp(cx - \int_{-\infty}^x (x-y)^3
\exp(cy - \tilde{q}(y)) dy )|
$$
can be estimated as
$$|
\exp(cx - \int_{-\infty}^x (x-y)^3
\exp(cy - \tilde{g}(y)) dy )
-
\exp(cx - \int_{-\infty}^x (x-y)^3
\exp(cy - \tilde{q}(y)) dy )|
\leq
$$
$$\exp(cz_{21r})|
\exp( - \int_{-\infty}^x (x-y)^3
\exp(cy - \tilde{g}(y)) dy )
-
\exp( - \int_{-\infty}^x (x-y)^3
\exp(cy - \tilde{q}(y)) dy )|
$$

Since both $$\int_{-\infty}^x (x-y)^3
\exp(cy - \tilde{q}(y)) dy>0 $$
and
$$\int_{-\infty}^x (x-y)^3
\exp(cy - \tilde{q}(y)) dy >0
$$ are positive
then
$$|
\exp( - \int_{-\infty}^x (x-y)^3
\exp(cy - \tilde{g}(y)) dy )
-
\exp( - \int_{-\infty}^x (x-y)^3
\exp(cy - \tilde{q}(y)) dy )|
\leq
$$
$$|
 \int_{-\infty}^x (x-y)^3
\exp(cy - \tilde{g}(y)) dy
-
 \int_{-\infty}^x (x-y)^3
\exp(cy - \tilde{q}(y)) dy |
$$

The last function allows the estimate
$$|
 \int_{-\infty}^x (x-y)^3
\exp(cy - \tilde{g}(y)) dy
-
 \int_{-\infty}^x (x-y)^3
\exp(cy - \tilde{q}(y)) dy |
\leq
$$
$$|
 \int_{-\infty}^x (z_{21r}-z_{in})^3
[\exp(cy - \tilde{g}(y))
-
\exp(cy - \tilde{q}(y))] dy |
$$
Since $g$ and $q$ are positive we come
to
$$|
 \int_{-\infty}^x (z_{21r}-z_{in})^3
[\exp(cy - \tilde{g}(y))
-
\exp(cy - \tilde{q}(y))] dy |
\leq
$$
$$(z_{21r}-z_{in})^3 \exp(cz_{21r})
|
 \int_{-\infty}^x
[ \tilde{g}(y)
-
 \tilde{q}(y)] dy |
$$
The final  inequality will be
$$
\int_{-\infty}^x
[ \tilde{g}(y)
-
 \tilde{q}(y)] dy
 \leq
 || q-g || (x-z_{in})
 $$

As the result we come to
$$
| g - q | \leq
(z_{21r} - z_{in})^3 \exp(cz_{21r})
|| g -q ||
(z -z_{in})^3 \int_{z_{in}}^z
(x-z_{in}) dx
$$
Having marked that
$(z -z_{in}) <(z_{21r} -z_{in})$
and
$(x-z_{in}) <(z_{21r} -z_{in})$ we see
that
$$
|| g - q || \leq
(z_{21r} - z_{in})^7 \exp(cz_{21r})
|| g -q ||(z - z_{in})
$$

Now we can formally organize
sequential application of the last
estimate
and having applied the last relation
$i$ times with explicit integration of
$(z - z_{in})$ we get
$$
|| g - q || \leq
(z_{21r} - z_{in})^7 \exp(cz_{21r})
|| g -q ||(z - z_{in})^i /i!
$$

One can easily see that the last r.h.s.
is the term in the Taylor's expansion
of
$$
(z_{21r} - z_{in})^7 \exp(cz_{21r})
|| g -q ||\exp(z - z_{in})
$$ with all standard consequences
mentioned above in such a situation.

The uniqueness for $z<z_{21r}$ is
proven.

In principle it is sufficient for all
finite $z_{21r}$. But below more fine
estimates  will be proven.

\subsection{Uniqueness for the big values
of arguments}

Here we shall use the function
$\xi$ or $\varphi$ instead of $cx - g$.
Again we mark by "$\tilde{}$" the solution
referred to the iterated equation.

To prove uniqueness for $z>z_{21r}$ one
can note that
$$
\exp(cx - \int_{-\infty}^x (x-y)^3
\exp(\tilde{\xi}) dy ) \leq
\exp(\xi_2) < max \exp(\xi_2)
$$

Then it is clear that
$$
\exp(cx - \int_{-\infty}^x (x-y)^3
\exp(\tilde{\xi}) dy )
\leq
R |cx - \int_{-\infty}^x (x-y)^3
\exp(\tilde{\xi}) dy|
$$
with some fixed constant $R$.
This already solves all problems. But
we shall present  a detailed derivation.

One can also note that the previous
consideration will be valid not only for
$z_{21r}$ but for any finite $z$. For
such a value $z_{21r}$ we shall choose
$z$ or $x$ at which already
$$
cx - \int_{-\infty}^x (x-y)^3
\exp(\tilde{\xi}) dy < 0
$$
Since one can prove that
$cx - \int_{-\infty}^x (x-y)^3
\exp(\tilde{\xi}) dy$ goes to $-\infty$
at big $z$ the last requirement could
be satisfied.

Let us suppose that there are two
different solutions.
For two different solutions $\tilde{\xi}$ and
$\tilde{\varphi}$
one can write
$$
|\tilde{\varphi} - \tilde{\xi}| \leq
\int_{-\infty}^z (z-x)^3
| (-cx+\int_{-\infty}^x (x-y)^3
\exp(\tilde{\varphi}) dy) -
$$
$$
 (-cx+\int_{-\infty}^x (x-y)^3
\exp(\tilde{\xi}) dy) | dx
$$

We can rewrite the last equation as
$$
|\tilde{\varphi} - \tilde{\xi}| \leq
\int_{-\infty}^z (z-x)^3
\int_{-\infty}^x (x-y)^3
| \exp(\tilde{\varphi})  -
\exp(\tilde{\xi})| dy  dx
$$

Since at $z<z_{21r}$ the solution is
unique then $\tilde{\varphi} =
\tilde{\xi}$
and we cut the lower limit of
integration up to $z_{21r}$. Then
$$
|\tilde{\varphi} - \tilde{\xi}| \leq
\int_{z_{21r}}^z (z-x)^3
\int_{z_{21r}}^x (x-y)^3
| \exp(\tilde{\varphi})  -
\exp(\tilde{\xi})| dy  dx
$$

Since it is known that both
$\tilde{\varphi}$ and $\tilde{\xi}$
have only one maximum and decrease at
$|z| \rightarrow \infty$ it is clear
that both $\tilde{\varphi}$ and
$\tilde{\xi}$ are less than some
constant and then
$$
| \exp(\tilde{\varphi})  -
\exp(\tilde{\xi})| \leq
Q | \tilde{\varphi}  -
\tilde{\xi}|
$$
with some
fixed\footnote{The value of $Q$ is close to $1$.}
constant $Q$.
Then
$$
|\tilde{\varphi} - \tilde{\xi}| \leq
Q \int_{z_{21r}}^z (z-x)^3
\int_{z_{21r}}^x (x-y)^3
| \tilde{\varphi}  -
\tilde{\xi}| dy  dx
$$
and
$$
||\tilde{\varphi} - \tilde{\xi}|| \leq
Q \int_{z_{21r}}^z (z-x)^3
\int_{z_{21r}}^x (x-y)^3
 dy  dx
 || \tilde{\varphi}  -
\tilde{\xi}||
$$

The chain of inequalities can be
prolonged
$$
||\tilde{\varphi} - \tilde{\xi}|| \leq
Q \int_{z_{21r}}^z (z-x)^3
\int_{z_{21r}}^x (z-y)^3
 dy  dx
 || \tilde{\varphi}  -
\tilde{\xi}||
$$
The r.h.s. can be rewritten in a
following way
$$
||\tilde{\varphi} - \tilde{\xi}|| \leq
Q (\int_{z_{21r}}^z (z-x)^3 dx)^2
 || \tilde{\varphi}  -
\tilde{\xi}||
$$
Then since $(z-x)< (z-z_{21r})$ one can
see that
$$
||\tilde{\varphi} - \tilde{\xi}|| \leq
Q (\int_{z_{21r}}^z (z-z_{21r})^3 dx)^2
 || \tilde{\varphi}  -
\tilde{\xi}||
$$

In other regimes of vapor consumption
it is important to prove the uniqueness
for the arbitrary positive power
$\alpha$ instead of $3$. So, we
get the following inequality
$$
||\tilde{\varphi} - \tilde{\xi}|| \leq
Q (\int_{z_{21r}}^z (z-z_{21r})^{\alpha} dx)^2
 || \tilde{\varphi}  -
\tilde{\xi}||
$$

Having calculated the integral we come
to
$$
||\tilde{\varphi} - \tilde{\xi}|| \leq
Q (z-z_{21r})^{2\alpha+2}
 || \tilde{\varphi}  -
\tilde{\xi}||
$$

Now we apply this inequality
$i$ times
with explicit integration
of $(z-z_{21r})$ and the
result of this integration.
Finally we  get
$$
||\tilde{\varphi} - \tilde{\xi}|| \leq
RQ \frac{(z-z_{21r})^{W(i)}}{V(i)}
 || \tilde{\varphi}_0  -
\tilde{\xi}_0||
$$
where
$\tilde{\varphi}_0$ and
$\tilde{\xi}_0$ are initial
approximations. For functions $V$ and
$W$ one can get some estimates
$$
V(i) > i!
$$
$$
W(i) < (2 [\alpha] +2 )i + const_1
$$
where $[\alpha]$ is the minimal integer
number greater than $\alpha$ and
$const_1$ is some constant which is
independent on $i$.

Now we can make the following remark:
We can assume that earlier we proved the
uniqueness until $z_{21r}+1$. Then
$z-z_{21r}> 1$.

Then we come to the following estimate
$$
||\tilde{\varphi} - \tilde{\xi}|| \leq
const Q \frac{(z-z_{21r})^{(2[\alpha]+2)i}}{i!}
(z-z_{21r})^{const_1}
$$

One can see that the r.h.s. is the
member of the Taylor's serial for the
function
$$
Q
const(z-z_{21r})^{const_1}
\exp((z-z_{21r})^{(2[\alpha]+2)})
$$
So, then the conclusion about the
uniqueness and existence of solution
can be also proven.

\section{$f$-representation }

One can rewrite the equation (\ref{1})
by the substitution
$$
x \rightarrow cx
$$
$$
z \rightarrow cz
$$
in a following form
$$
g(z) = f \int_{-\infty}^z (z-x)^3
\exp(x-g(x)) dx
$$
with $f=c^{-4}$.

This representation is more convenient
to show the uniqueness of dependence of
$z_{max}$ on $c$.

Suppose that the are two  parameters
$f_1$ and $f_2$. Let  it be $f_2 >
f_1$.

Then
$$
g_{f_1} (z) = f_1 \int_{-\infty}^z
(z-x)^3 \exp(x-g_{f_1}(x)) dx
$$
$$
g_{f_2} (z) = f_2 \int_{-\infty}^z
(z-x)^3 \exp(x-g_{f_2}(x)) dx
$$

For $g_{f_2}$ we construct iterations
with initial approximation
$$
g_{f_2\ 0} (x) = g_{f_1} (x)
$$
and a recurrent procedure
$$
g_{f_2\ i+1} (z) = f_2 \int_{-\infty}^z
(z-x)^3 \exp(x-g_{f_2\ i}(x)) dx
$$

Then
$$
g_{f_2\ 1} (z) =f_2 \int_{-\infty}^z
(z-x)^3 \exp(x-g_{f_1}(x)) dx
= (f_2/f_1) g_{f_1}(x)
$$
So,
$$
g_{f_2\ 1} (z)
> g_{f_1}(x) (1+\epsilon)
$$
with finite positive small $\epsilon$.

One can see the following important
fact

{\bf
The following  formula
$$
\frac{d|g_{f_2\ 1}-g_{f_2\ 0}|}{dz} > 0
$$
takes place.
}

This  is clear if one notes that
$$
(f_2 - f_1)
\int_{-\infty}^z (z-x)^3
\exp(x-g_{f_1\ i}(x)) dx
$$
grows when $z$ grows.

Really
$$
3 (f_2 - f_1)
\int_{-\infty}^z (z-x)^2
\exp(x-g_{f_1\ i}(x)) dx
>0
$$
This proves the inequality.

One can also prove
the following statement:

{\bf
There exists a point $z`_{in}$
until which $g_{f2\ 2} > g_{f2\ 0}$
}

This follows from the explicit
estimates for the first and the second
iterations with zero initial
approximations. It is known that at $z
\rightarrow - \infty$ they come very
close and estimate the solution very
precise. Really,
$$
g_{f_1} \rightarrow g_{f_1\ as} \sim
 6 f_1 \frac{\exp(cx)}{c^4} = 6 f_1
 \exp(x)
$$
$$
g_{f_2} \rightarrow g_{f_2\ as} \sim
 6 f_2 \frac{\exp(cx)}{c^4}
 = 6 f_2 \exp(x)
$$

Then
$$
g_{f_2 \ 2 \ as} = g_{f_2 \ 1\ as }
= g_{f_2 \ as}
$$
Since
$$
g_{f_2 \ 1\ as } > g_{f_2\ 0} = g_{f_1}
$$
we see that asymptotically
$$
g_{f_2\ 2} > g_{f_2 \ 0}
$$

Then the necessary point exists.
This proves the statement.

As the result we see that at extremely
big negative $z$ the following hierarchy takes
place
$$
g_{f2\ 0} \leq g_{f2\ 2} \leq ... \leq g_{f2\ 3}
\leq  g_{f2\ 1}
$$

So, all of them are between
$g_{f2\ 0}$ and $g_{f2\ 1}$
at extremely big negative $z$.
Later iterations can go away from this
frames.  We shall mark the coordinate
when iteration number $i$
attains the mentioned boundary by
$z`_i$.

One can easily see that either $z`_i$
goes to infinity or $z`_i$ exists. One
can also see that
$$
z`_{i+1} \geq z`_i
$$

It is easy to see because the
integration is going only for the
values of argument smaller than the
current one. This proves the estimate.

A question appears whether the limit
$$\lim_{i\rightarrow \infty} z`_i$$
exists.

The existence of a limit means that at
this limiting point (let it be
$z`_{\infty}$) the limit of odd
iterations differs from the limit of
even iterations.  But both the limit of
odd iterations and the limit of even
iterations belong to the solutions of
the
iterated equation. But we have already
proved that the solution of the iterated
equation is the unique one. So, the
contradiction is evident. We come to
the
conclusion that there is no such a
limit.

The last conclusion particularly means
that
$$
g_{f_2} > g_{f_1}
$$

Really, the analogous reasons show that
the possible points of crossing
$g_{f_1\ 2j+2}$, $g_{f_1\ 2j+4}$, etc. with
 $g_{f_1\ 2j}$ for every $j$
 cannot have a finite
 limit. So, for every finite $z$
 we see that
 $$
 g_{f_2} >g_{f_2\ 2j}
 $$

Since for every finite $z$
there is such $j$ which provides
$$
g_{f_2\ 2j} > g_{f_2\ 0} = g_{f_1}
$$
we see that
$$
g_{f_2} > g_{f_1}
$$
This proves the necessary inequality.

The last inequality leads to the
important relation
$$
\frac{dg_f}{df} > 0
$$
for any given value of the argument.

\section{The uniqueness of the root}

We analyze the question: how many
solutions with different $c$ (or $f$)
can have the maximum of $cx-g$
(or $x-g_f$)
at $x=0$?

The most evident answer is that there
will be only one solution but this has
to be proven. To prove it we shall use
$f$-representation.

The equation for the coordinate
$z_{max}$ of the spectrum will be the
following
$$
1 = 3 f \int_{-\infty}^{z_{max}}
(z_{max} - x)^2 \exp(x-g(x)) dx
$$
Then
$$
f = \frac{1}{3}
(
\int_{-\infty}^{z_{max}}
(z_{max} - x)^2 \exp(x-g(x)) dx
)^{-1}
$$

Now we can calculate
$ df/dz_{max} $ Then we shall take it
at $z_{max} = 0$ and if we are able to
prove that $df/dz_{max} < 0$ then the
necessary property will be established.

The possibility to take here $z_{max} =
0$ is ensured by existence of the root
which was proven a few sections
earlier. Now we see that the proof of existence
of the root   was
really necessary.

For the derivative $df/dz_{max}$ one
can get the following equation
$$
\frac{df}{dz_{max}} =
- \frac{1}{3}
(
\int_{-\infty}^{z_{max}}
(z_{max} - x)^2 \exp(x-g(x)) dx
)^{-2}
Y
$$
where
$$
Y =
\frac{d\int_{-\infty}^{z_{max}}
(z_{max} - x)^2 \exp(x-g(x))
dx}{dz_{max}}
$$

The direct calculation gives
$$
Y = 2 \int_{-\infty}^{z_{max}} (z_{max}
- x)
\exp(x - g(x)) dx -
\int_{-\infty}^{z_{max}} (z_{max}
- x)^2
\exp(x - g(x)) \frac{dg}{df} dx
$$
In the last relation one has to
consider $\frac{dg}{df}$ as a function
of $x$.

One can also get for $Y(0)$ the
following equation
$$
Y(0) = \frac{1}{3}
\frac{d^2 g}{dz^2}|_{z=0}
 - \int_{-\infty}^{0}  x^2
\exp(x - g(x)) \frac{dg}{df} dx
$$

Now we shall calculate
$dg/df$. One can get the following
equation
$$
\frac{dg}{df}
=
\int_{-\infty}^x (x-y)^3
\exp(y-g(y)) dy -
f\int_{-\infty}^x (x-y)^3
\exp(y-g(y)) \frac{dg}{df} dy
$$
One can rewrite the last relation as
$$
\frac{dg}{df}
=
g/f -
f\int_{-\infty}^x (x-y)^3
\exp(y-g(y)) \frac{dg}{df} dy
$$
Since
$dg/df >0$ (it has been already proven)
then
the sub-integral function
$(x-y)^3
\exp(y-g(y)) \frac{dg}{df}$
is positive and the we have
$$
f\int_{-\infty}^x (x-y)^3
\exp(y-g(y)) \frac{dg}{df} dy \geq 0
$$
Then
$$
0 \leq
\frac{dg}{df}
\leq
g/f
$$

Now we shall return to the function
$Y$.
One can rewrite it as
$$
Y(z_{max}) =
\int_{-\infty}^{z_{max}}
(z_{max} - x)
[ 2 - (z_{max} - x) \frac{dg}{df} ]
\exp(x-g(x)) dx
$$
Then
$$
Y(0) =
\int_{-\infty}^{0}
(0 - x)
[ 2 - (0 - x) \frac{dg}{df} ]
\exp(x-g(x)) dx
$$

The sign of $Y(0)$ is important.

Then we consider the function
$$
l =[ 2 + x \frac{dg}{df} ]
$$
We have to recall that $x$ takes here
only negative values.

Then
$$
l > [ 2 + x \frac{g}{f} ]
$$

For $g$ one can take the estimate
$$
g < f \exp(x)
$$
going from the first iteration with
initial zero approximation.
Then
$$
l > [ 2 + x \exp(x) ]
$$

One can easily see that the
function $x\exp(x)$ has
only one minimum at $x=-1$ where this
function is $-\exp(-1)$.

Then
$$
l> 2 - exp(-1) >0
$$

Then $Y(0) >0$ and
$$
df/dz_{max}|_{z_{max}=0} < 0
$$

It means that the uniqueness of the
root is proven.

\section{Uniqueness of solution of the
problem B}

{\bf
Suppose that $\xi_1$ and $\xi_2$ are
two different solutions of the problem
B. Then (this is the statement)
$$
\int_{-\infty}^0 y^2 \exp(-\xi_1(y))
dy
=
\int_{-\infty}^0 y^2 \exp(-\xi_2(y))
dy
$$
}

We shall prove this equality.

Suppose that these integrals do not
equal, let it be
$$
\int_{-\infty}^0 y^2 \exp(-\xi_1(y))
dy
>
\int_{-\infty}^0 y^2 \exp(-\xi_2(y))
dy
$$
Then for the problem A
one can see that
$c_1>c_2$
and
$f_1 < f_2$.
Then (because there is only one root of
equation $z_{max}(f) = 0$) we come to the
conclusion that at
least one solution can not have maximum
of $x-g_f(x)$ at $z=0$. So, at least
one solution is not a solution of the
problem B.

We come to the contradiction.

This completes the proof.

We can formulate another statement:

{\bf The solution of the
problem B is unique. }

It is clear because as it is proven
$$
\int_{-\infty}^0 y^2 \exp(-\xi_2(y))
dy
$$
has one fixed value. Then it is the
problem of solution of equation
(\ref{1}) with fixed $c$. The last
problem has the unique solution.

This completes the justification of the
uniqueness of solution of the problem
B.

{\bf
One can also see that the solution of the
problem A is the solution of the
problem B.
}

To see this
one can simply differentiate the
solution and then use the condition for
derivative.

Since the solution of the problem A is
unique and solution of the problem B is
unique one can easily see that:

{\bf The
solution of the problem A and the
problem B coincide.}

\section{Similarity of the forms of
the size distributions}

Having established the uniqueness of
solution of equation (\ref{1}) with
given
$c$ (or $f$), the uniqueness
of the  solution of the problem A and
the uniqueness of the solution of the
problem B we can
investigate the correspondence between
solutions with different $c$.

The main result will be the following:

{\bf
Actually all
solutions of equation (\ref{1})
 with different $c$ (or $f$)
are  identical.
}

Really, in equation (\ref{1}) we
fulfill
the substitution
$$zc \rightarrow z$$
$$xc \rightarrow x$$
and come to
$$
g(z) = \frac{1}{c^4}
\int_{-\infty}^z (z-x)^3 \exp(x - g(x))
dx
$$

The next transformation will be the
shift
$$z  \rightarrow z+ \Delta$$
$$x  \rightarrow x+ \Delta$$
We choose
$$
\Delta \sim  4\ln c
$$

The factor $(z-x)$
can not change. the region of
integration will be actually the same -
from infinity up to a "current moment" $z$.

Then
the amplitude
$1/c^4$ can be easily canceled.

So, the spectrum, i.e. the function
$$
\Phi = \exp(x-g)
$$
has one and same form, which differs
only by rescaling of
the argument and the
rescaling of the amplitude.
The numerical simulations confirm this
conclusion (see the figure 2)

\begin{figure}[hgh]

\includegraphics[angle=270,totalheight=8cm]{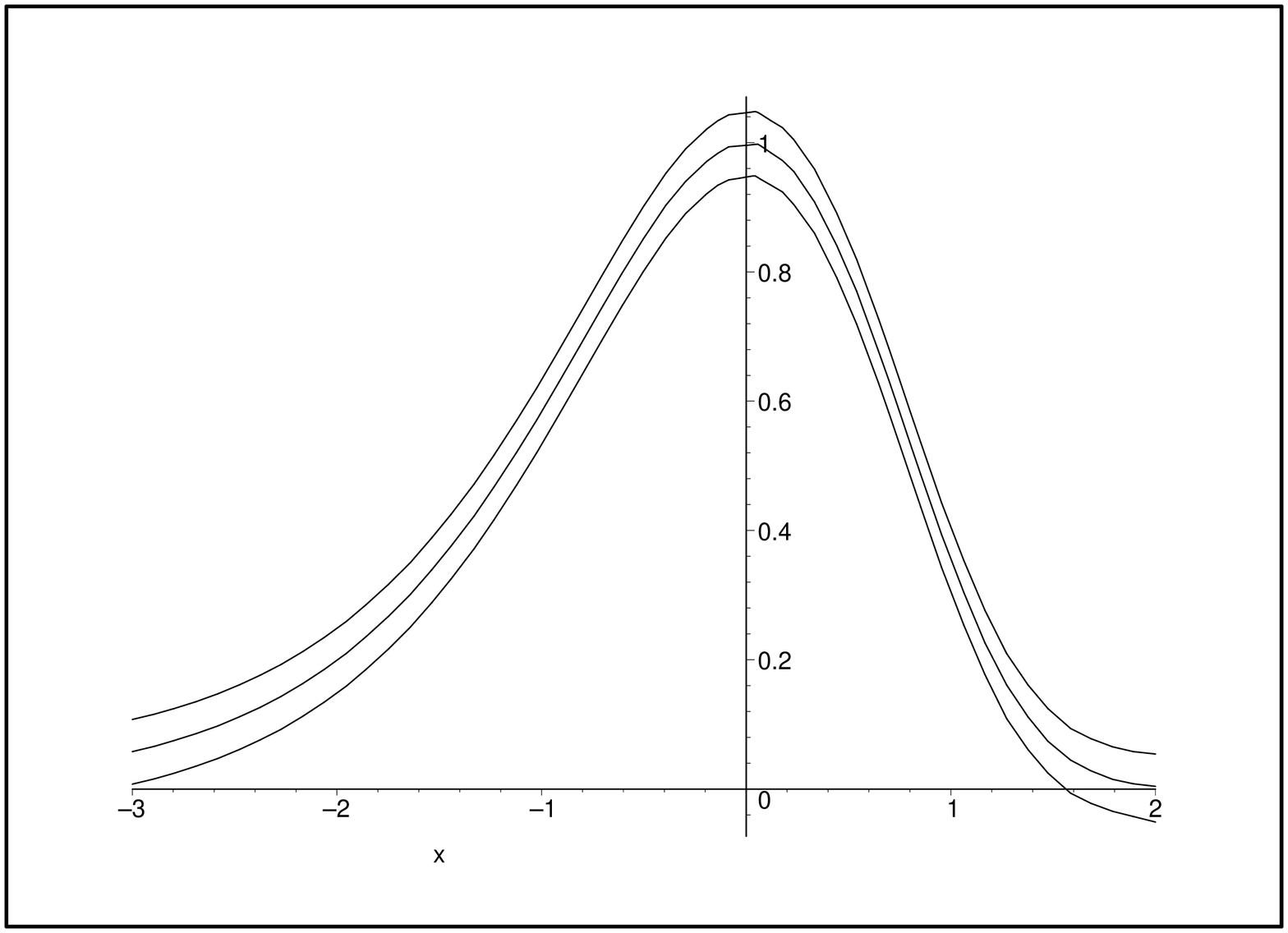}

\begin{caption}
{  Forms of distributions (with the shifts -0.05; 0; +0.05 for
$c=0.1; 1; 10$}
\end{caption}

\end{figure}

The shift in these pictures was introduced
only in order to
see three curves. When the shift will
be zero then only one curve can be
seen. The coincidence is practically
ideal.

The certain question whether there can
be simply multiple repeating of
solutions appears here. But the
uniqueness of solution of equation
(\ref{1}) with given $c$ and the
property
$dg/df >0$ established earlier
ensure the absence of such
repeating.

\end{document}